\documentclass{nature_incl_fig}
\usepackage{lineno}
\usepackage{graphicx}
\usepackage{caption}
\usepackage{amsmath}
\usepackage[utf8]{inputenc}
\usepackage[T1]{fontenc}
\usepackage{sansmath,amssymb, amsmath, graphicx, grffile}

\usepackage{nameref} 

\usepackage[dvipsnames]{xcolor}

\usepackage[margin=1in]{geometry}
\usepackage{xspace}
\usepackage[colorlinks=true,citecolor=blue,linkcolor=blue,urlcolor=blue,breaklinks,hypertexnames=false]{hyperref} 

\begin{document}

\renewcommand\baselinestretch{1.2}

\begin{center}
    {\Large \bf The observed impact of galaxy halo gas on fast radio bursts}
\end{center}
\vspace{0.25cm}
\begin{center}
    {\large Liam Connor$^{1,2,*}$ \& Vikram Ravi$^{1,2}$} \\
\end{center}

\vspace{0.3cm}

\noindent $^{1}$ Cahill Center for Astronomy and Astrophysics, MC\,249-17, California Institute of Technology, Pasadena CA 91125, USA. \\
$^{2}$ Owens Valley Radio Observatory, California Institute of Technology, Big Pine CA 93513, USA. \\
$^{*}$ E-mail: liam.dean.connor@gmail.com \\

\clearpage

{\bf  Galaxies and groups of galaxies exist in dark-matter halos filled with diffuse gas. The diffuse gas represents up to 80\% of the mass in baryonic matter within the halos\cite{anderson-bregman-2010, tumlinson-2017}, but is difficult to detect because of its low density (particle number densities of $\lesssim10^{-4}$\,cm$^{-3}$) and high temperature (mostly greater than $10^{6}$\,K). 
Here we analyze the impact of diffuse gas associated with nearby galaxies using the dispersion measures (DMs) of extragalactic fast radio bursts (FRBs). FRB DMs provide direct measurements of the total ionized-gas contents along their sightlines. 
Out of a sample of 474 distant FRBs 
from the CHIME/FRB Catalog 1\cite{chime-frbcat1}, we identify a subset of events that likely intersect the dark-matter halos of galaxies in the local Universe ($<40$\,Mpc).
The mean DM of the galaxy-intersecting FRBs is larger than 
the non-intersecting DMs with probability $>0.99$ and the excess DM 
is $>90$\,pc\,cm$^{-3}$ with $>95\%$ confidence.
The excess is larger than expected for the diffuse gas surrounding isolated galaxies, but may be explained by additional contributions from gas surrounding galaxy groups, including from the Local Group. 
This result demonstrates the predicted ability of FRBs to be used as sensitive, model-independent measures of the diffuse-gas contents of dark-matter halos\cite{mcquinn-2014, ravi-2019, prochaska-zheng-2019,ksz19}}.

The Canadian Hydrogen Intensity Mapping Experiment (CHIME) Fast Radio Burst Project (CHIME/FRB hereafter) published its
first catalog of 535 FRBs detected in the 400--800\,MHz radio band between 2018 July 25 and 2019 July 1\cite{chime-frbcat1}. CHIME is a transit telescope that observes the whole northern sky daily\cite{bandura-2014, chime-kaspi-2017}. The first CHIME/FRB catalog includes one-off or ``non-repeating'' bursts from 474 sources, and 18 repeating sources that have been detected two or more times. 
The repeaters are not included in our analysis (Methods). This is the largest collection of FRBs from a single survey by more than an order of magnitude. The uniformity 
of the sample, both in sky position and in terms of selection effects, provides significant power for statistical inference\cite{chime-frbcat1}. For example, a correlation has been observed between the locations of CHIME FRBs and the distribution of known galaxies with cosmological redshifts $0.3<z<0.5$, which indicates that an order-one fraction of CHIME FRBs originate from correspondingly large distances\cite{Rafiei-2021}.  

Despite its central role in the formation and evolution of galaxies, diffuse gas surrounding galaxies is notoriously difficult to study in detail\cite{tumlinson-2017}. We refer to this medium as the circumgalactic medium (CGM) in the case of individual galaxies (dark matter halo masses $10^{11}M_{\odot}\lesssim M_{h}\lesssim10^{12.5}M_{\odot}$), the intra-group medium (IGrM) in the case of galaxy groups ($10^{12.5}M_{\odot}\lesssim M_{h}\lesssim10^{13.5}M_{\odot}$), and the intracluster medium (ICM) in the case of galaxy clusters ($M_{h}\gtrsim10^{13.5}M_{\odot}$). As short-duration (approximately millisecond) radio-wavelength pulses, FRBs are dispersed during propagation through astrophysical plasmas along their sightlines. The DM to a redshift $z$ is given by the following integral\cite{Shull-danforth-2018}:
\begin{equation}
    \mathrm{DM}(z) = \frac{c}{H_{0}}\int_{0}^{z}\frac{(1+z)n_{e}(z)}{\sqrt{(1+z)^{3}\Omega_{M} + \Omega_{\Lambda}}}dz,
\end{equation}
where $c$ is the vacuum speed of light, $H_{0}$ is the Hubble constant, $n_{e}(z)$ is the free-electron number density as a function of redshift, $\Omega_{M}$ is the fractional cosmic matter density, and $\Omega_{\Lambda}$ is the fractional cosmic dark-energy density. The observed DM, ${\rm DM}_{\rm obs}$, is conventionally broken down into a series of components:
\begin{equation}
    \rm DM_{\mathrm{obs}} = DM_{\mathrm{MW}} + DM_{\mathrm{MW halo}} + \sum^{N_{gal}}_i DM_{CGM_i} + DM_{\mathrm{IGM}} +  DM_{\mathrm{host}},
\end{equation}
where ${\rm DM}_{\mathrm{MW}}$ and ${\rm DM}_{\mathrm{MW halo}}$ are contributed by the Milky Way interstellar medium and gaseous halo respectively, ${\rm DM}_{\rm host}$ is contributed by the FRB host galaxy, ${\rm DM}_{\rm IGM}$ is contributed by the intergalactic medium, and the CGM contributions from intervening galaxies ($\rm DM_{CGM_i}$) are summed over $N_{\rm gal}$ objects. The intergalactic-medium contribution, DM$_{\rm IGM}$, is often the largest term and is thought 
to lead to the Macquart (DM--$z$) relation\cite{Macquart-2020}. The exact value of $\rm DM_{CGM_i}$ for a given intervening galaxy will depend on the impact parameter of the background FRB ($b_{\perp}$), and the unknown spatial gas-density distribution surrounding the galaxy. Individual FRBs have been 
found to intersect foreground halos\cite{prochaska-askap-2019, connor-2020-m33}, but without the statistics of a larger sample it is difficult to extract the DM contribution of the CGM\cite{mcquinn-2014, ravi-2019, ksz19}.

CHIME is a compact telescope and lacks the angular resolution to unambiguously identify the host galaxies of FRBs\cite{Eftekhari-2018, Aggarwal-PATH-21}. Typical sky-localization error regions are $\sim$\,0.2$\times$0.2\,degrees$^2$ in the CHIME/FRB Catalog 1. We therefore focus this work on nearby, massive galaxies with dark-matter halo virial radii that subtend large angular scales. An approximately Milky-Way sized halo with a virial radius of $\sim$\,200\,kpc will be larger in angular size than a standard CHIME/FRB localization region at distances $\lesssim$\,40\,Mpc. 

We identify a sample of CHIME FRBs that likely intersect nearby halos within a threshold impact parameter (Methods). We have utilized the Gravitational Wave Galaxy Catalogue (GWGC) containing the position, distance, and $B$-band absolute magnitude ($M_B$) of 53,255 galaxies within 100\,Mpc\cite{gwgc-2010}. GWGC is likely complete for all galaxies within 40\,Mpc with $M_B<-18$; we note that our results below are only likely to be diluted by any incompleteness. We begin by using only galaxies with $M_B<-18.5$, that are within 40\,Mpc, and are more distant than 500\,kpc. For a fiducial threshold impact parameter of $b_\perp=200$\,kpc, we find 26 intersecting FRBs. We consider it an intersection if the impact parameter between a CHIME FRB and GWGC galaxy is less than the threshold value, $b_\perp$, scaled by galaxy mass (see Methods). In Figure 1, we show the sky locations of the full CHIME/FRB catalog of non-repeating FRB sources, as well as the GWGC galaxy / FRB intersectiond within this fiducial $b_{\perp}$. The list of CHIME/FRB galaxy intersections is presented in Table 1. 

The subset of FRBs that intersect the halos of nearby galaxies have statistically higher extragalactic DMs than the rest of the CHIME/FRB sample. 
We calculate the extragalactic DM by subtracting a model for the ${\rm DM}_{\rm MW}$ along each FRB sightline\cite{ne2001}. In this fiducial sample, we exclude FRBs observed at Galactic latitudes within $\pm$\,5\,degrees of the Galactic plane to mitigate systematic errors in the model for ${\rm DM}_{\rm MW}$. To determine the significance of the mean-DM excess, we employ two statistical tests (see Methods for details). 
The first is a jackknife re-sampling of the data, 
corresponding to a nonparametric test that does not make 
any assumptions about the underlying DM distribution. 
Randomized FRBs are put through the same cross-matching procedure with $\lesssim$\,40\,Mpc GWGC galaxies as the real data, and the ``intersecting'' CHIME/FRB sources are compared with the ``non-intersecting'' FRB DMs. This allows 
us to ask, in what fraction of random samples do we see a DM excess larger than the true data. This is a one-sided test. The second statistic we use is a two-sample one-sided t-test, 
which is designed explicitly to look for a positive shift in the 
mean DM of galaxy-intersecting FRBs. 

For both statistical tests, we analyze the significance of the observed DM excess as a function of $b_\perp$. In Figure 2, we show the DM excess for different choices of $b_\perp$ together with the results from 1000 jackknife re-samplings of the data. 
For $b_\perp$ between 75\,kpc and 300\,kpc, we find the jackknife and one-sided t-test p-values are all less than $5\%$ and in only $0.5\%$ of the resampled curves are the mean DM excesses as extreme as those of the real, unshuffled data (Methods). We have shown through simulation that neither test is biased towards small p-values and both are more sensitive to shifts in mean DM than a more generic nonparametric test, the Kolmogorov–Smirnov (KS) test (see Methods and Extended Data Figure 1). 

As a further test of robustness, we apply different cuts to the CHIME/FRB catalog and GWGC. In Figure 3, we plot the Galactic latitudes and extragalactic DMs of the intersecting and non-intersecting CHIME FRBs. We find no correlation between extragalactic DM and Galactic latitude for both sub-samples of CHIME FRBs (see Methods). This shows that the observed excess DM of the sample of intersecting FRBs is unlikely to be due to errors in modeling the Milky Way disk and halo DM contributions. For the CHIME/FRB sources, we apply various limits on the Galactic latitude (between 0\,deg and 20\,deg from the Galactic plane), and consider an additional model for ${\rm DM}_{\rm MW}$\cite{ymw16} together with no subtraction of ${\rm DM}_{\rm MW}$.  We also alter the $M_{B}$ cut on foreground galaxies between $-18.0$ and $-19.5$, and vary the maximum galaxy distance between 15\,Mpc and 50\,Mpc. In all of these cases the excess DM persists. However, if we consider only GWGC galaxies at distances between 50\,Mpc and 100\,Mpc the signal disappears, likely because the halos subtend small solid angles relative to the CHIME/FRB localization regions, and the apparent intersections are false. We consider additional means of identifying CHIME/FRB galaxy intersections, and find no change to our results (Methods). 

It is unlikely that the FRBs that we associate with nearby galaxy halos in fact originate from these galaxies (Methods). We also find that the total number of intersections is consistent with the expected range for a background population, based on the probability of halo intercepts (Methods). It is also unlikely that these FRBs pass through the disks of the nearby galaxies, both from a statistical perspective, and because the effects of multi-path propagation would render them undetectable given the CHIME observing frequencies\cite{pn18}. The excess DM in the CHIME/FRB galaxy intersections can therefore be robustly ascribed to gas in the halos of the nearby galaxies (including the Local Group galaxies) because there is no other difference in the sample selection. 

We next compute the observed DM excess over a range of impact parameters. For the fiducial sample of CHIME/FRB galaxy intersections ($b_{\perp}=200$\,kpc), we find a mean DM of 790$\pm$110\,pc\,cm$^{-3}$, and the remainder of the CHIME/FRB sample has a mean DM of 600$\pm$20\,pc\,cm$^{-3}$ ($1\sigma$ errors throughout). We note that this is just one choice of $b_{\perp}$ and there is more statistical power in the full DM excess curve. The fiducial value histogram and estimated DM excess for different $b_\perp$ are shown in Figure 2. For $b_\perp<300$\,kpc, the DM excess is $\gtrsim75$\,pc\,cm\,$^{-3}$. This does not necessarily mean that the diffuse gas extends to 300\,kpc. Each data point is highly correlated between threshold impact parameters, so excess DM at smaller $b_\perp$ will appear at larger $b_\perp$ as well. Not all the CHIME/FRB galaxy associations are secure, and positional uncertainty will act to bias the measured DM excess at fixed $b_{\perp}$ downward.

The DM excess ($>90$\,pc\,cm$^{-3}$ at $>95\%$ confidence for $75\,$kpc$<b_\perp<300$\,kpc; see Methods)
is substantially larger than predicted for diffuse gas in the halos of individual galaxies\cite{mcquinn-2014, prochaska-zheng-2019}. By estimating the halo masses of the sample of intersected galaxies listed in Table 1, we find predicted DM excesses less than 40\,pc\,cm$^{-3}$ even in the case that these halos retain the expected cosmic fraction of baryons (see Methods and Extended Data Figure 2). However, over two thirds of the FRB/galaxy intersections are with galaxies that belong to groups (Table 1); this is not unexpected\cite{yang07}. Intersected groups include the Local Group, the M81 Group, and the M74 Group. Galaxy groups host a rich IGrM, aspects of which have been extensively studied through extended thermal X-ray emission\cite{mulch2000}, ultraviolet absorption spectroscopy\cite{stocke14}, and its impact on the cosmic microwave background (the Sunyaev-Zel'dovich / SZ effect)\cite{planck-2013}. It is likely that galaxy groups retain significant fractions of their expected baryon contents in a hot ($>10^{6}$\,K) phase, although the effects of energy feedback may complicate the exact distributions of temperature and density\cite{Oppenheimer-2021}. The observed DM excess is nonetheless consistent with simple models for the DM contribution from the IGrM (Methods)\cite{fujita17,prochaska-zheng-2019}.  

If we consider only the CHIME FRBs that intersect the halos of Local Group galaxies M31 and M33, we see excess DM, but it is not statistically significant on its own (jackknife p-value\,$\sim$\,0.08 at the fiducial $b_\perp$). 
If the Local Group IGrM is modeled with a dark matter halo mass of $10^{12.5}\,M_\odot$, it could contribute significant DM over tens of degrees on the sky in the direction of M31\cite{prochaska-zheng-2019}. 
The Local Group IGrM has recently been studied with analyses of X-ray O\,VII and O\,VIII emission lines and SZ maps that indicate significant hot-gas content extended to an angular radius of approximately $30$\,degrees from M31\cite{Qu-2021}. The authors suggested this represents a hot-gas bridge connecting the Milky Way with M31 and M33 that corresponds to a DM of 80--400\,pc\,cm$^{-3}$. We anticipate that our measurements, when augmented by additional data from CHIME/FRB and other instruments, will significantly aid in the modeling and interpretation of the X-ray and SZ observations of the Local Group IGrM, together with the IGrM of other nearby intersected groups. 

The magnitude of the observed DM excess in the CHIME/FRB galaxy intersections is promising for future FRB-DM measurements of the mass and density profile of diffuse gas surrounding galaxies and galaxy groups. For example, with a larger FRB sample, the analysis presented here can be applied to galaxy clusters, where the 10$^{14-15}\,\rm M_\odot$ halos with virial radii of around 2\,Mpc may contribute DMs greater than $10^3$\,pc\,cm$^{-3}$\cite{fujita17,prochaska-zheng-2019}. We note that the most highly dispersed CHIME/FRB source, with extragalactic $\rm DM=3006.7$\,pc\,cm$^{-3}$, appears to intersect within 1.4\,Mpc of the center of the galaxy cluster J185.83917+56.47005 at $z$=0.328, found by cross-matching with the GMBCG cluster catalog\cite{GMBCG-2013}.

\vspace{1cm}
\newpage
\section*{References}
\bibliographystyle{naturemag-doi}
\bibliography{biblio_all}

\noindent {\bf Supplementary Information} is linked to the online version of the paper at \\ www.nature.com/nature.

\noindent {\bf Acknowledgements.}
We first the referees, whose careful 
consideration and suggestions were invaluable. 
We thank Cameron Hummels, Wenbin Lu, J. Michael Shull, and the Caltech FRB group for a helpful discussions. We also 
thank Calvin Leung, Kiyoshi Masui, and Mohit Bhardwaj for 
valuable comments on the manuscript. 
This  research  was  partially  supported  by  the  National  Science Foundation  under grant AST-1836018.

\noindent {\bf Author contributions.}
V.R. conceived of searching only nearby foreground galaxies for FRB/halo interceptions. L.C. developed the methods for cross-matching the catalogs, statistically testing the DM distributions, and analyzing the DM excess that are reported in the main text figures and results. V.R. modeled the halo DM contribution shown in Extended Data Figure 3. L.C led the writing of the manuscript in close collaboration with V.R.


\noindent {\bf Competing interests statement.} The authors declare that they have no competing financial interests.

\clearpage

\begin{figure}

\centering
\includegraphics[width=\textwidth]{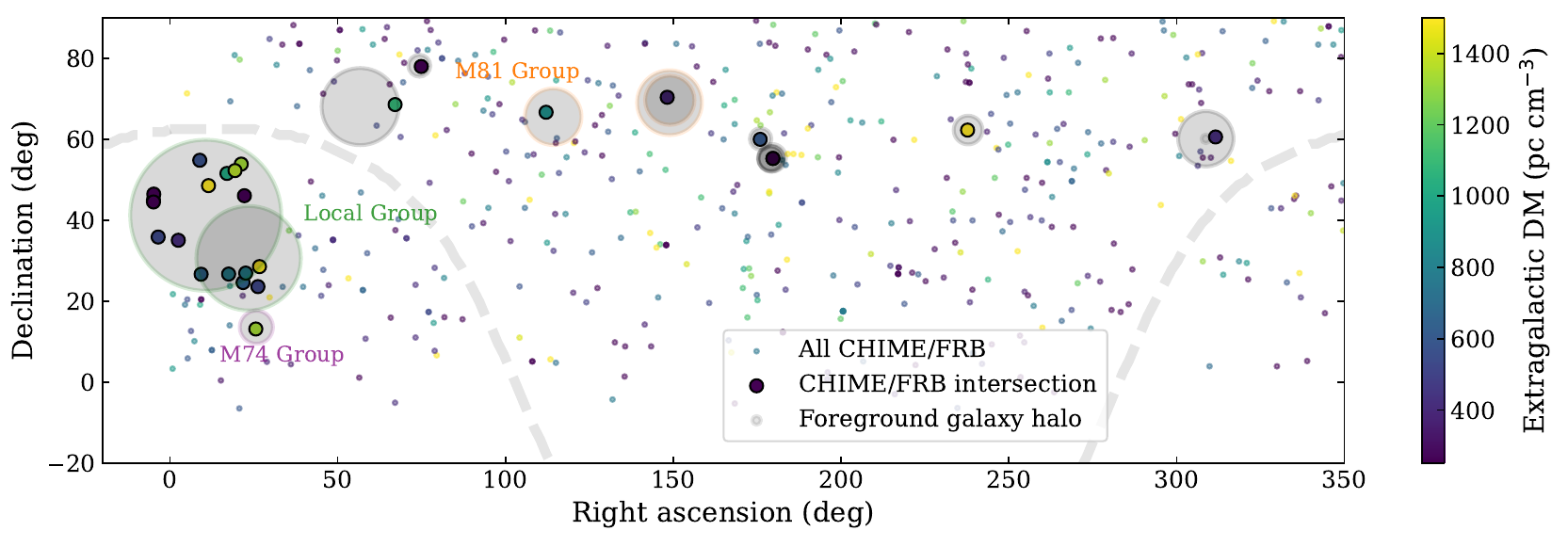}
\caption{{\bf CHIME/FRB galaxy intersections.} We show the sky locations of 474 one-off FRB sources in the CHIME/FRB Catalog 1\cite{chime-frbcat1}, with those that intersect nearby ($<40$\,Mpc) GWGC galaxies (grey circles) using the fiducial $b_\perp=200$\,kpc indicated by larger circles. The color of FRBs encodes their individual extragalactic DMs (as indicated in the color bar).
Although the color bar is saturated at 1500\,pc\,cm$^{-3}$, some FRBs that intersect foreground galaxy halos have significantly higher DMs (Table 1). The radii of the filled grey circles correspond to the angular size using the galaxy's estimated virial radius at that distance. Some specific galaxy groups are indicated by colored circles around the galaxy locations. The Galactic 
plane is shown as a light dashed line.
    \label{fig:1}}
\end{figure}

\clearpage


\begin{figure}
\centering
    \includegraphics[width=0.7\textwidth]{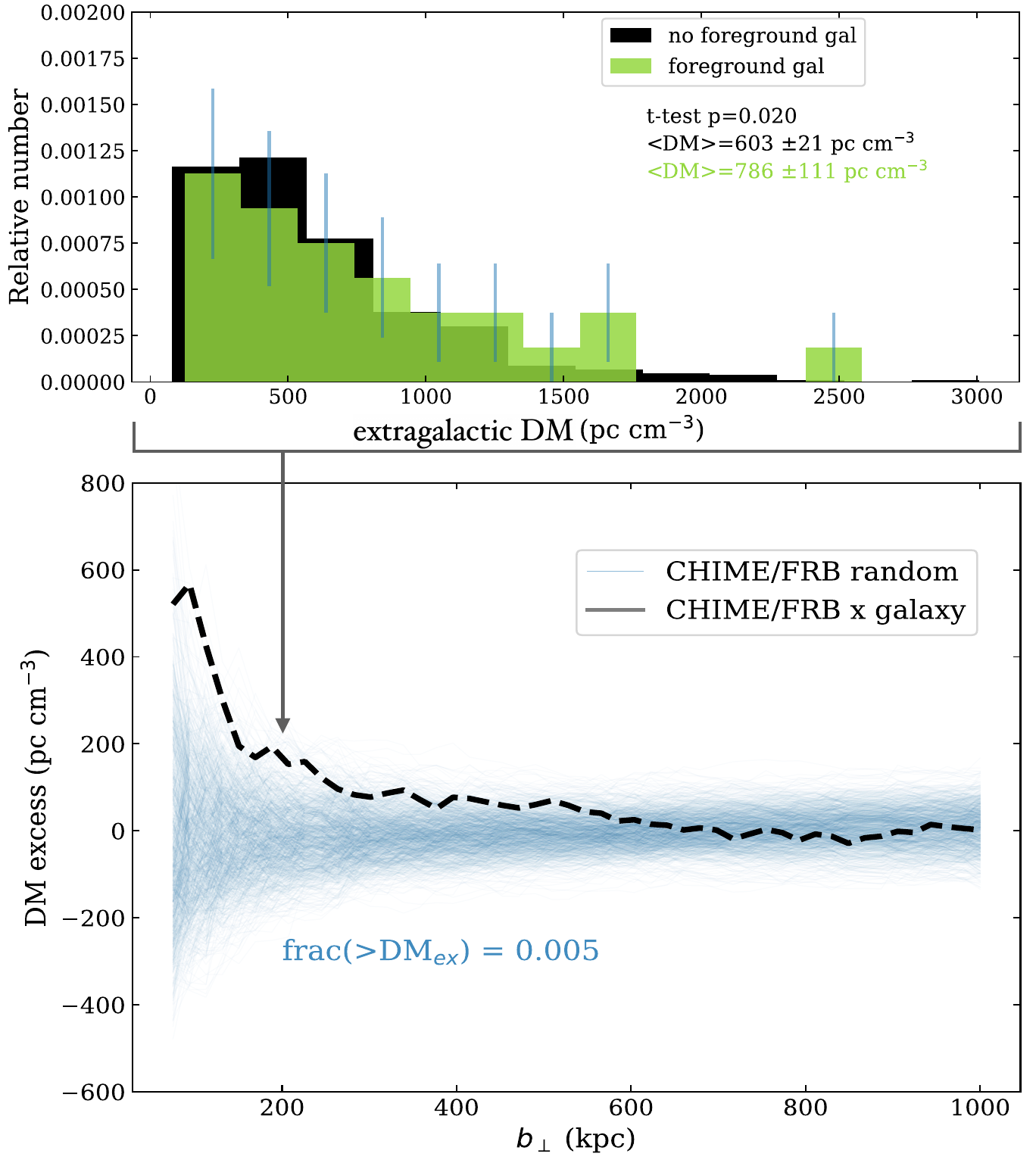}\\
    \caption{{\bf Statistical significance of the 
    excess DM.} The top panel shows normalized extragalactic DM distributions of CHIME FRBs 
    that intersect foreground galaxies (green) and those 
    that do not (black) for a single fiducial value 
    for the threshold impact parameter of $b_\perp=200$\,kpc, with $1\,\sigma$ Poissonian error bars. The vertical axis on the bottom panel shows  
    excess DM as a function of the chosen threshold impact parameter $b_\perp$ for the full range. The excess is
    the difference in mean DMs between galaxy-intersecting 
    and non-intersecting FRBs (black dashed curve). A jackknife test 
    with a random shuffling of DMs among the CHIME/FRB sources (thin blue curves) 
    is also shown for 
    1000 realizations. In just 0.5$\%$ of cases, the mean excess DMs 
    of the random jackknife realizations are greater than the mean measured 
    excess averaged between 75\,kpc and 300\,kpc.
    \label{fig:3}}
\end{figure}

\clearpage

\begin{figure}
    \centering
    \includegraphics[width=0.8\textwidth]{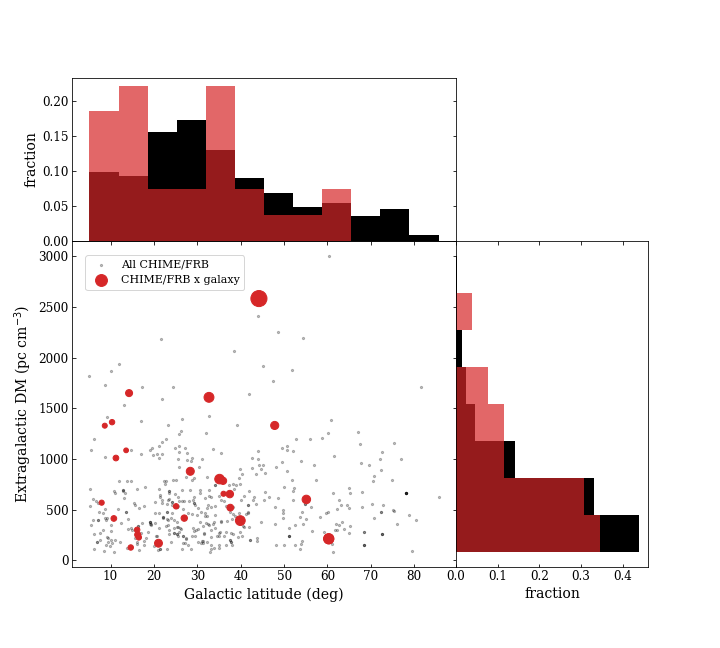} \\
    \caption{{\bf The extragalactic DM and Galactic latitude 
    distributions of CHIME/FRBs in our sample.} 
    The FRBs that intersect foreground galaxies 
    are shown in red (for $b_{\perp}=200$\,kpc) and the non-intersecting sources are shown in black.
    In the scatter plot, marker size for the 
    galaxy-intersecting FRBs is inversely proportional 
    to their impact parameter with the foreground galaxy. 
    There is no apparent correlation between Galactic latitude 
    and extragalactic DM, in either of the FRB subsets (Methods).
    The top and right panels correspond to fractional histograms 
    for the two subsets of CHIME/FRB sources. The characteristically larger DM of intersecting FRBs is evident in the histogram in the right panel.}
    \label{fig:2}
\end{figure}

\clearpage

\begin{table*}
	\centering
	\captionsetup{labelformat=empty}
	\caption{{\bf Table 1:} A table of FRB/foreground galaxy intersections 
	         using CHIME/FRB Catalog 1 and GWGC for galaxies within 
	         40\,Mpc. We have 
	         held $b_\perp$ constant at 200\,kpc. Group identifications and masses are from a recent catalog of groups with radial velocities of $<3500$\,km\,s$^{-1}$, within which all intersected galaxies are found.\cite{kourkchi17}}
	\label{tab:e1}
	\begin{tabular}{ccccccc} 
		\hline
		{\bf FRB name} & {\bf Galaxy} & {\bf Group} & {\bf $\log$ $M_{\rm vir}/M_{\odot}$} & {\bf Dist (Mpc)} & {\bf $R_\perp$ (kpc)} & {\bf DM (pc cm$^{-3}$)}\\
		\hline
FRB20190223B & NGC6946 & NGC6946 & 12.1 & 5.9 & 153 & 411.8 \\ 
FRB20190430B & NGC6015 & None & & 15.2 & 19 & 2583.0 \\ 
FRB20190423A & NGC3998 & NGC3998 & 12.6 & 14.1 & 48 & 211.0 \\ 
FRB20181214C & NGC3835 & None & & 34.1 & 69 & 599.5 \\ 
FRB20190612A & NGC3034 & M81 & 12.6 & 3.9 & 52 & 390.6 \\ 
FRB20190701D & NGC2403 & M81 & 12.6 & 3.2 & 78 & 877.4 \\ 
FRB20180925A & IC0391 & None & & 20.8 & 80 & 167.1 \\ 
FRB20190211A & IC0342 & Maffei & 12.7 & 3.3 & 221 & 1084.4 \\ 
FRB20190130A & NGC0660 & M74 & 12.2 & 9.2 & 76 & 1330.1 \\ 
FRB20190128A & M33 & LG & 12.4 & 0.8 & 90 & 652.5 \\ 
FRB20190217B & M33 & LG & 12.4 & 0.8 & 54 & 800.4 \\ 
FRB20190226C & M33 & LG & 12.4 & 0.8 & 95 & 783.3 \\ 
FRB20190605D & M33 & LG & 12.4 & 0.8 & 51 & 1607.7 \\ 
FRB20190607A & M33 & LG & 12.4 & 0.8 & 108 & 518.5 \\ 
FRB20181018A & M31 & LG & 12.4 & 0.8 & 154 & 1008.5 \\ 
FRB20181101A & M31 & LG & 12.4 & 0.8 & 200 & 1327.8 \\ 
FRB20181129A & M31 & LG & 12.4 & 0.8 & 165 & 299.2 \\ 
FRB20181130A & M31 & LG & 12.4 & 0.8 & 170 & 125.0 \\ 
FRB20181224A & M31 & LG & 12.4 & 0.8 & 163 & 225.2 \\ 
FRB20190102A & M31 & LG & 12.4 & 0.8 & 202 & 655.6 \\ 
FRB20190106A & M31 & LG & 12.4 & 0.8 & 132 & 251.2 \\ 
FRB20190116E & M31 & LG & 12.4 & 0.8 & 173 & 1362.6 \\ 
FRB20190122B & M31 & LG & 12.4 & 0.8 & 123 & 415.1 \\ 
FRB20190415B & M31 & LG & 12.4 & 0.8 & 188 & 567.6 \\ 
FRB20190614C & M31 & LG & 12.4 & 0.8 & 170 & 531.7 \\ 
FRB20190628C & M31 & LG & 12.4 & 0.8 & 101 & 1649.6 \\ 
		\hline
	\end{tabular}
\end{table*}

\clearpage

\begin{center}
    {\Large \bf Methods}
\end{center}
\vspace{0.5cm}

{\bf Sample selection.} We used the 474 non-repeating CHIME/FRB sources published in Catalog 1\cite{chime-frbcat1}. We exclude the 18 repeating sources in case they have different distance or DM distributions than apparent non-repeaters. This could arise from the bias towards seeing two or more bursts from nearby objects\cite{lu-2019, frbpoppy-2019}, or if repeaters have preferentially larger ${\rm DM}_{\rm host}$. The CHIME/FRB Catalog 1 has three different DM fields: `dm\_fitb' is the best fit total DM of the observed FRB. `dm\_exc\_ne2001' and `dm\_exc\_ymw16' correspond to total DM minus the expected Milky Way contribution in that direction from the NE2001\cite{ne2001} and YMW16\cite{ymw16} models respectively. We are concerned here with extragalactic DM, so we opt to use `dm\_exc\_ne2001'. However, all three fields produce similar results in our analysis. To avoid potential uncertainties in modeling the Galactic interstellar medium, we exclude CHIME/FRB sources with an absolute value of Galactic latitude less than 5\,degrees, and consider a range of limits on the 
maximum absolute value between 0 and 20 degrees. 
We do not attempt to subtract 
the Milky Way halo because its hot gas is assumed to be isotropic and would not affect our excess DM statistic \cite{keating-cgm}. 

The Gravitational Wave Galaxy Catalogue (GWGC) contains sky position, distance, $B$-band magnitude, and galaxy type for 53,255 galaxies within 100\,Mpc\cite{gwgc-2010}. We select only galaxies at distances less than 40\,Mpc such that their angular size is large enough to associate with the CHIME/FRB uncertainty region. We also set a minimum distance of 0.5\,Mpc so as to exclude globular clusters and the Magellanic Clouds. Finally, we use only galaxies whose absolute $B$-band magnitude is less than $-18.5$, and explore limits between $-18.0$ and $-19.5$. The fiducial cut leaves 2829 galaxies. This corresponds to a density of roughly 0.01\,Mpc$^{-3}$. The optical depth of galaxies with an average virial radius $R_{vir}$ out to a maximum distance $d_{max}$ is then $\tau\approx0.01\,\mathrm{Mpc}^{-3}\cdot\, d_{max} \cdot\pi R^2_{vir}\approx 0.05$ if we assume the mean 
virial radius is 200\,kpc and $d_{max}$ of 40\,Mpc. We therefore expect approximately 20--25 CHIME/FRB-galaxy intersections. This is in line with our observations.

We have cross-matched the CHIME/FRB Catalog 1 with a subset of the GWGC to look for FRBs for which,
\begin{equation}
    R_\perp(\theta_{ML}) \lesssim R_{vir}
\end{equation}
which we parameterize as,
\begin{equation}
    R_\perp(\theta_{ML}) \leq b_\perp \times \left(\frac{\mathrm{M}_h}{1.3\times10^{12} \mathrm{M}_\odot}\right)^{1/3}.
\end{equation}
\noindent Here, $R_\perp$ is the impact parameter of the CHIME/FRB source at the reported maximum likelihood position, $\theta_{ML}$. M$_h$ is the dark matter halo mass estimated from the galaxy's absolute $B$-band magnitude and $b_\perp$ is the impact parameter threshold for a galaxy with a Milky Way halo mass. We allow $b_\perp$ to vary for Figure 2, but fix it at 200\,kpc for Figures 1 and 3. We point out that at low $b_\perp$ the number of galaxy-intersecting FRBs becomes small and the variance on excess DM is large. This is one reason we choose to compare the excess DM averaged over a range of effective impact parameters. We note that the DM excess tends towards zero for large impact parameter, as expected. We have tried two other methods for associating CHIME FRBs with galaxies. The first is to ignore galaxy mass and assume a halo was intersected if $R_\perp(\theta_{ML}) \leq b_\perp$. The second was to use the CHIME/FRB localization confidence interval arrays, provided with their data release\footnote{https://chime-frb-open-data.github.io/localization.html}. Using these arrays, we take a confidence-weighted average of the impact parameter as opposed to using a single maximum-likelihood position. If this weighted mean $\left < R_\perp(\theta_{ML}) \right >$ is less than some value, it is considered an interception. In all three cross-matching methods, we find similar results for the statistical significance of the difference in mean DMs, as well as similar amplitudes for the DM excess. We have also tried including the DM selection function 
reported by CHIME/FRB\cite{chime-frbcat1} in order to account for 
incompleteness. We fit their DM recall curve 
with a polynomial and divide the observed DM distributions for the
intersecting and non-intersecting subsets by that curve. We then re-compute 
p-values on the de-biased DM distributions. We find 
the statistics are not significantly changed, which is in line with expectation as CHIME is relatively complete in DM and both subsets of data have the same selection effects.

\textbf{Statistical significance.}
Our analysis focuses on two primary statistical tests for 
DM excess: a jackknife resampling of the CHIME/FRB data and 
Student's one-sided t-test. The former is a nonparametric order statistic 
in that it makes no assumptions about the underlying DM 
distributions. 
Its ``p-value'' is the fraction of resamplings 
in which the DM excess is larger than the true data.
To compute this, we first calculate the 
excess DM as a function of $b_\perp$ of the real data (see Figure~2). The jackknife resampling 
is done by randomly shuffling the CHIME/FRB DM data while keeping the source positions constant, and computing excess 
DM at each threshold impact parameter. This 
is done for 10,000 realizations, of which 1000 are shown in Figure 2. We then ask, how many 
of those resampled DM/$b_\perp$ curves have a larger 
sum than the real data in the range 75--300\,kpc. 
We find that only 0.5$\%$ of resamplings have 
a larger excess DM between galaxy-intersecting 
and non-intersecting FRBs when compared with the 
real data. Changing the range over which the excess 
is calculated from 75--300\,kpc to other reasonable values 
does not significantly affect our result. 75\,kpc is 
our lower limit because of the small number of intersections 
below that impact parameters, but the effect remains 
if that number is set to 0\,kpc. Similarly, 300\,kpc is 
about double the typical virial radius of Milky Way-like 
galaxies, and allows for the inclusion of more massive 
halos, such as galaxy groups.

The p-value of our one-sided two-sample t-test can be interpreted as the
probability that the difference in mean DMs between 
the two subsets of FRBs would be at least as extreme as what is 
found in the data, if there were no positive DM excess. 
Typical p-values are below 0.05 for the expected virial 
radii of the foreground galaxies and the average p-values 
at $b_\perp \lesssim 300$\,kpc are smaller than in more 
than 99$\%$ of resamplings in our jackknife test.
However, the t-test is a parametric test, meaning it makes certain assumptions about the 
underlying data. First, it expects that the means of the two samples being compared are normally distributed---a consequence of the central limit theorem. And in the case 
of Student's one-sided t-test, the variance of the two samples ought to be roughly equal. Nonetheless, we point out that t-tests are often robust against moderate 
deviations from the two aforementioned assumptions\cite{bland1995}. We believe the DMs of CHIME/FRB sources can be appropriately compared with Student's t-test.

To test this claim explicitly, we have simulated CHIME/FRB data with a similar DM distribution as Catalog 1 in order to test the validity of the one-sided t-test. We model the CHIME/FRB DM distribution 
using a Gamma distribution with shape parameter, $\alpha=2$,  
scale parameter, $\theta_\Gamma = 40$, with the DMs scaled by a factor of 8 to match the empirical distribution. For each 
realization in this simulation, we draw 474 FRB DMs from the 
Gamma distribution and randomly select 25 as having 
intersected a ``galaxy''. We then apply a Student's t-test 
to the galaxy-intersecting sample of FRBs and the remaining 
449 simulated sources. 
The p-values are uniformly distributed between 0 and 1 and $\sim$5$\%$ of realizations have $p<0.05$, despite the 
DM data being non-Gaussian and the sample sizes to be different 
by over an order of magnitude. We conclude that the parametric nature of the t-test we have used on the real data is not a major concern. If it were, we would have expected a non-uniform distribution of p-values and more than 5$\%$ of realizations to have $p<0.05$, which would indicate an artificially inflated rate of low p-values.

The jackknife and t-test are only two inferential 
statistics that we could have applied to the data. 
One common choice for a nonparametric statistic  
comparing two samples is the Kolmogorov–Smirnov (KS) test. The KS-test does not lead to a highly significant detection when applied to our data.
However, the KS-test is non-ideal for our purposes because 
it is a generic test to determine if two samples 
are drawn from the same distribution. We are physically-motivated 
to search for the more specific signal of a positive shift in mean DM 
in FRBs that intersect foreground galaxies, for which both our particular jackknife and Student's one-sided t-test are designed.
The KS-test, on the other hand, has less 
statistical power to look for shifts in median or mean because 
it tests more for deviations from the null hypothesis\cite{lehmann1998nonparametrics}. In Extended Data Figure 1 we compare three methods for computing p-values on our simulated CHIME/FRB DM data both in the case where there is no excess DM (top row) and the case where excess 
DM is added to galaxy-intersecting sources (bottom row). The excess 
is normally distributed with mean 150\,pc\,cm$^{-3}$ and standard 
deviation 50\,pc\,cm$^{-3}$. We find that the t-test and jackknife test are more sensitive to shifts in the mean DM 
than the KS-test.

A scenario where we expect the DM excess signal to disappear is when the foreground galaxies are too far away for the CHIME/FRB localization region to be unambiguously matched with the galaxy's halo. While there will be many ``intersections'' due to the increasing optical depth with distance, most will be false positives because the CHIME/FRB localization uncertainty becomes significantly larger than the halo for distances beyond 40\,Mpc. We therefore do not expect excess DM if the minimum galaxy distance is set to $\gtrsim$\,40\,Mpc in the GWGC catalog. And indeed, we find only p-values greater than 0.05 with this cut, despite dozens of reported intersections at 100\,kpc\,$\leq b_\perp \leq 300$\,kpc.

\textbf{Quantifying the DM excess.} We can quantify the excess DM by taking an average in the 
range 75\,kpc\,$\leq b_\perp \leq 300$\,kpc. The 1\,$\sigma$ error is estimated from the distribution of 
DM excesses in 10,000 jackknife tests.
This gives a 90$\%$ confidence 
interval of 90--420\,pc\,cm$^{-3}$ on the excess DM. However, the upper bound is not robust. It may be that not all  FRB/galaxy associations included here are correct because of CHIME localization uncertainties and the uncertain extents of halos. This would lead to our sample being contaminated by non-intersecting FRBs. Further, the impact parameters that we assign to the FRB/galaxy associations are uncertain for the same reasons. As more sky area is covered by larger $b_{\perp}$ values than by smaller $b_{\perp}$, it is likely that the true impact parameters of the sample of FRB/galaxy intersections are larger. This would also bias low our estimate of the DM excess at a specific $b_{\perp}$. We therefore only regard the lower bound as robust.

\textbf{Interpretation: gas in galaxy halos.} Our understanding of the warm/hot gaseous contents of dark matter halos hosting galaxies, galaxy groups and galaxy clusters remains foggy. However, it has long been known that this medium is of significant astrophysical importance, because it represents between 80--90\% of the baryon contents of dark matter halos\cite{anderson-bregman-2010, shull-2012, tumlinson-2017}. Fundamental open questions include the total mass and radial density profile of the CGM/IGrM/ICM; FRB observations are a promising new technique to address these questions\cite{mcquinn-2014, ravi-2019, prochaska-zheng-2019, keating-cgm, ksz19}. We express the CGM/IGrM/ICM mass as a fraction, $f_{\rm gas}$, of the expected baryonic mass in a given dark-matter halo, $\Omega_{b}M_{h}/\Omega_{M}$, where $\Omega_{b}=0.049$ is the fractional cosmic baryon density and $\Omega_{M}=0.315$ is the fractional cosmic matter density\cite{planck-2018}.

Existing observational probes of $f_{\rm gas}$ for individual galaxies and galaxy groups are beset by systematic issues, and yield uncertain results in comparison with galaxy cluster measurements. The thermal X-ray emission and thermal/kinetic Sunyaev-Zel'dovich (SZ) measurements that are used for galaxy clusters become less reliable for lower halo masses because of decreasing emission columns, lower gas pressures, and the increased effects of feedback from astrophysical processes in galaxies. Individual galaxy halos at low redshifts retain a multi-phase CGM, with a high covering fraction of cool ($\sim10^{4}-10^{5}$\,K; and therefore primarily ionized) clouds interspersed within the hot ($\gtrsim10^{6}$\,K), diffuse gas probed by X-ray and SZ observations\cite{tumlinson-2017, bregman-2018}. Intriguingly, stacked SZ observations of galaxies and galaxy clusters are beginning to reveal a potentially self-similar relation between the inferred signal and $M_{h}$, indicating both a universal $f_{\rm gas}$ close to unity and radial density profile\cite{planck-2013,lmw+20}. This is consistent with extrapolations of X-ray observations of the IGrM\cite{mulch2000}, but potentially in tension with X-ray observations of individual galaxies\cite{bregman-2018}. The tension may be resolved by considering feedback effects in different samples, and the existence of significant gas fractions at sub-virial temperatures. Significant other uncertainties remain, including the mass fractions of cool gas in galaxy halos\cite{tumlinson-2017}, and indeed the true extents of galaxies\cite{shull14}. 

In the following, we consider the expected DM excess for our sample of FRB sightlines that intercept nearby galaxy halos. We first consider each galaxy in isolation, and derive the expected DM accrued for each intercept under two models for the radial number-density distribution of baryons in halos, $n(r)$: 
\begin{itemize}
    \item The fiducial modified Navarro-Frenk-White (mNFW) profile described in\cite{prochaska-zheng-2019}.
    \item A `beta' model\cite{jf84,mb15}  
    \begin{equation}
        n(r) = n_{0}\left(1+\left(\frac{r}{r_{c}}\right)^{2}\right)^{-3\beta/2},
    \end{equation}
    where $n_{0}$ is a normalization constant, $r_{c}$ is a core radius that we fix to 1\% of the virial radius, and $\beta$ is an index that we set to 0.5 for consistency with observations of galaxy and galaxy-cluster halos\cite{bregman-2018}.
\end{itemize} 
We fix $f_{\rm gas}=1$, and approximately derive the halo masses by scaling the approximate Milky Way halo mass\cite{watkins19} of $1.3\times10^{12}M_{\odot}$ by the ratio between the absolute $B$-band magnitudes of the GWGC galaxies and the Milky Way $B$-band magnitude of $-20.8$\cite{kkh+04}. We account for the primordial helium fraction in deriving the DM for each intercept. 

The distribution of expected DM excesses for the fiducial sample of 26 FRB / nearby-galaxy intercepts for $b_{\perp}<200$\,kpc is shown in Extended Data Figure 2. We find mean excess DMs for the mNFW and beta models of 34.7\,pc\,cm$^{-3}$ and 38.7\,pc\,cm$^{-3}$ respectively, well below the measured 95\% confidence lower bound of 90\,pc\,cm$^{-3}$. This analysis also demonstrates how the distribution of excesses is sensitive to the chosen radial density profile. More centrally concentrated profiles like the mNFW model result in distributions with longer tails. Consistent with previous works that derive the expected DM excesses for typical galaxy-halo intercepts\cite{ravi-2019, prochaska-zheng-2019}, we conclude that gas within individual galaxy halos is insufficient to explain the observed excess in the FRB / nearby-galaxy sample. Although this analysis is approximate by factors of a few due to uncertainties in the halo mass determination, the conclusions are robust because of our conservative assumption of $f_{\rm gas}=1$, and because of the downward bias discussed above. 

To illustrate how this problem may be solved, we next consider the expected excess DM accrued by FRBs intersecting halos of different masses. Using the same models for $n(r)$ as above, we simulate 100 samples of 26 random intercepts for local ($z=0$) halos of different masses. For each realization, we randomly draw 26 FRB positions using the uncertainties (assumed to parameterize elliptical Gaussian distributions) in right ascension and declination in the CHIME/FRB Catalog 1. We then calculate the projected offsets from the centres of the halos they intersect (as listed in Table 1). In calculating the predicted mean DM excess of the sample, an FRB is assigned a zero value if it intersects the halo beyond the virial radius. We do not use the confidence-level maps that accompany each Catalog 1 FRB because these cannot be directly translated into posterior predictive distributions on the FRB positions. We also emphasise that these simulations are illustrative rather than directly representative of reality, because we assume that all FRBs are intersecting halos of the same mass. The resulting mean DM excesses for halo masses $10^{12}M_{\odot}\leq M_{h}\leq10^{13.5}M_{\odot}$ are shown in Extended Data Figure 3. Two cases are considered:
\begin{itemize}

\item In the top panel of Extended Data Figure 3, we assume $f_{\rm gas}=1$, and consider the two models for the radial density distribution of baryons introduced above.
\item In the bottom panel of Extended Data Figure 3, we instead only consider the modified NFW model, but adopt results from four different cosmological simulations for $f_{\rm gas}$ at different halo masses\cite{Oppenheimer-2021}: SIMBA\cite{simba}, the IllustrisTNG 100\,Mpc box (TNG100)\cite{tng100}, EAGLE\cite{eagle}, and ROMULUS\cite{romulus}. The variations in $f_{\rm gas}$ between simulations are large, lying in the range 0.2--0.85, and the values typically increase by factors of a few from lower to higher halo masses. These variations are largely due to the (different) feedback prescriptions, which result in different amounts of gas being evacuated from halos. The simulation results are synthesized by Oppenheimer et al.\cite{Oppenheimer-2021}. Results for halo masses below $10^{12.5}M_{\odot}$ are not given, and we use the values at $10^{12.5}M_{\odot}$ for lower halo masses. 

\end{itemize}

Consistency between the simulated and measured DM excesses is only achieved for halo masses well in excess of the Milky Way halo mass. Consistent with previous results\cite{prochaska-zheng-2019}, this analysis makes it clear that membership of the intercepted galaxies in modest galaxy groups can explain the observed DM excess. The larger DM offset for the modified NFW profiles for larger halo masses (in the $f_{\rm gas}=1$ panel) is due to the fixed impact parameter that is assumed. For the predicted values of $f_{\rm gas}$, only some simulations (ROMULUS, and possibly EAGLE) are consistent with our observations, given the range of group virial masses listed in Table 1. However, several other uncertainties need to be accounted for in making a quantitative comparison, including uncertainties in the observed virial masses, the true values of $b_{\perp}$, the radial density profiles, and variance between the evolutionary histories of halos at a given mass. 


\textbf{Interpretation: excess host DM.} In this section, we consider and reject the possibility that the effect that we observe is explained by a high DM local to FRB host galaxies (${\rm DM}_{\rm host}$). This possibility relies on assuming that FRBs that appear to be spatially coincident with nearby galaxy halos in fact originate from within the nearby galaxies. It may then be possible that, by random chance, these nearby FRBs have substantially larger values of ${\rm DM}_{\rm host}$ than the remainder of the CHIME sample. Alternatively, if all FRBs have very large values of ${\rm DM}_{\rm host}$ in their rest frames, nearby FRBs will be observed to have larger values of ${\rm DM}_{\rm host}$ than more distant FRBs, because of the suppression of the local DM contribution by a factor $(1+z)^{-1}$. If all CHIME FRBs are observed at sufficiently low redshifts such that ${\rm DM}_{\rm host}$ is the dominant contributor to the total extragalactic DM, then it is possible that the extragalactic DMs of FRBs in fact reduce with increasing redshift. In this contrived example, sources that are close to nearby galaxies on sky will have higher observed DM than those that are not. 

The former scenario is unlikely given the large required ${\rm DM}_{\rm host}$ values of the local FRBs. The characteristic redshift of the CHIME sample of $0.3\lesssim z \lesssim 0.5$\cite{Rafiei-2021} implies a characteristic ${\rm DM}_{\rm IGM}\gtrsim300$\,pc\,cm$^{-3}$.\cite{Shull-danforth-2018} Assuming no significant evolution in FRB properties at redshifts $z\lesssim0.5$, the local sample would need to have  ${\rm DM}_{\rm host}$ values that are more than $\sim300$\,pc\,cm$^{-3}$ larger than those of the remainder of the CHIME sample. For reasonable distributions of ${\rm DM}_{\rm host}$ generated by considering different FRB progenitor scenarios\cite{walker20}, this is statistically unlikely. 

The latter scenario is also not supported, because characteristic values of ${\rm DM}_{\rm host}\gg300$\,pc\,cm$^{-3}$ would be required. This is inconsistent with the DMs of the majority of CHIME FRBs\cite{chime-frbcat1}. The extragalactic DMs of FRBs with host-galaxy redshift measurements typically increase with increasing redshift, and are dominated by gas external to galaxies\cite{Macquart-2020}. The distributions of fluence and DM for FRBs observed at the Parkes, ASKAP, FAST and CHIME telescopes are consistent with a population that exhibits a positive correlation between distance and DM\cite{shannon18,james21a,james21b,niu21,chime-frbcat1}. Most FRBs localized to host galaxies have ${\rm DM}_{\rm host}\lesssim300$\,pc\,cm$^{-3}$\cite{heintz20}. 
It is additionally possible that FRBs with high values of ${\rm DM}_{\rm host}$ will also undergo excess temporal broadening due to interstellar scattering in their host galaxies, which would make them harder to detect\cite{connor19,simard21}. 

There are two interesting counterpoints, but they cannot on their own produce our observed DM excess. The first is 
FRB\,190520, whose host DM could be as high as $10^3$\,pc\,cm$^{-3}$ \cite{190520}. The second comes from  cross-correlating galaxies between redshifts 0.3 and 0.5 with high-DM CHIME FRBs (DM$\gtrsim800$\,\,pc\,cm$^{-3}$) \cite{Rafiei-2021}. There is 
a significant cross-correlation signal between those two samples, indicating that some FRBs in that redshift 
range have large host galaxy dispersion ($\geq400$\,pc\,cm$^{-3}$). In fact, 
many CHIME FRBs with large extragalactic DM are at $z\sim0.4$, even though most $z\sim0.4$ 
FRBs do not have large extragalactic DM. 
These CHIME FRBs, however, are distributed 
randomly over two galaxy footprints (DESI and WISExSCOS) that are far more distant than 
our sample of foreground galaxies, which 
are at $z\approx0$; they do not preferentially intersect nearby galaxies. Finally, our 
galaxy-intersecting FRBs span a much wider 
DM range. Combining these facts, we find 
no evidence for the minority of FRBs with 
large host galaxy dispersion producing 
the DM excess reporting in this paper.

\noindent {\bf Data availability statement.} The data used in these analyses are all publicly available. The CHIME/FRB Catalog 1 can be found at \url{https://www.chime-frb.ca/catalog}. The Gravitational Wave Galaxy Catalogue (GWGC) can be downloaded online at \url{http://vizier.u-strasbg.fr/viz-bin/VizieR?-source=GWGC}.

\noindent {\bf Code availability statement.} This research made use of the open-source numpy, scipy, astropy, hmf, NFW,  CHIME/FRB's ``cfod'' package, and frb (\url{https://github.com/FRBs/FRB/tree/main/frb}) python packages. All custom code used in our analysis will be made available upon request. 

\clearpage

\clearpage

\begin{figure*}
     \centering
     \includegraphics[width=0.95\textwidth]{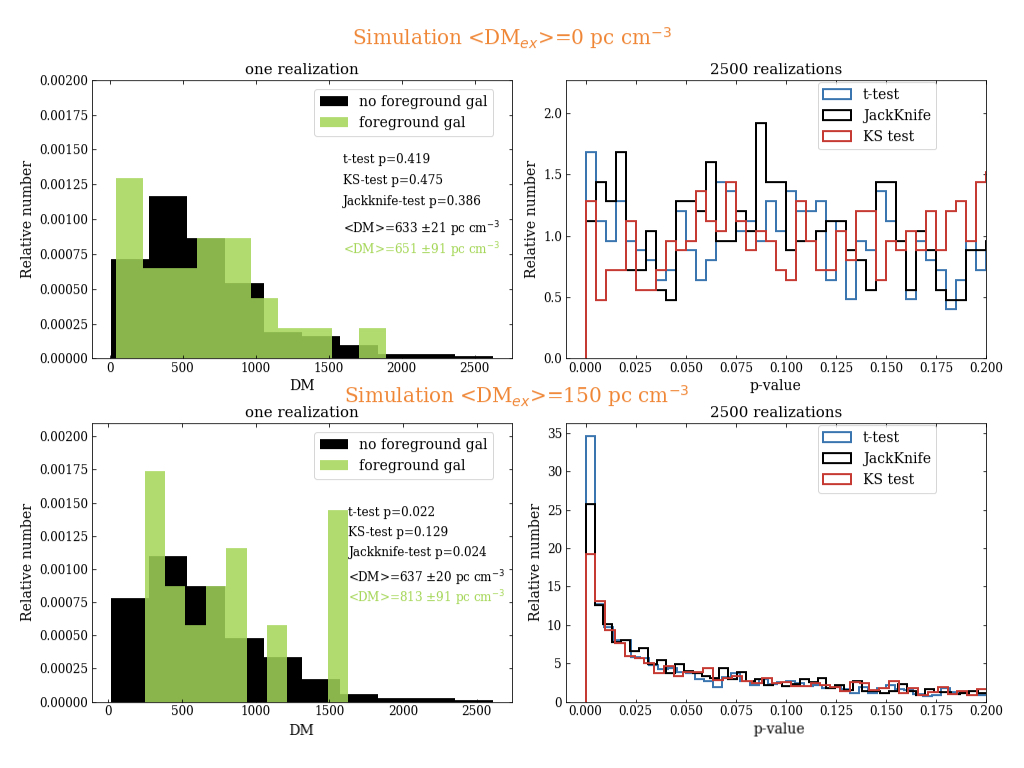}
     \captionsetup{labelformat=empty}
     \caption{{\bf Extended Data Figure 1: A simulation of 
     CHIME/FRB DMs to compare the statistical power and 
     appropriateness of three tests: Student's one-sided t-test, 
     our jackknife test, 
     and the Kolmogorov–Smirnov (KS) test.} We simulate 474 
     FRB DMs with a similar distribution to the CHIME/FRB 
     sources. In the top row, we have not added any excess 
     DM to the galaxy-intersecting sources. In the bottom row, 
     excess DM has been added to the 25 simulated FRBs that intersect 
     a foreground galaxy, with a normal distribution 
     of mean 150\,pc\,cm$^{-3}$ and 
     standard deviation 50\,pc\,cm$^{-3}$. From the top row, 
     we see that none of the tests produces spurious low p-values and their p-values are uniform 
     as expected. 
     The bottom right panel demonstrates that the KS-test is less
     sensitive to DM offsets than the t-test and the non-parametric jackknife test. The t-test and 
     jackknife tests are one-sided, in that they explicitly look for a positive mean DM difference, whereas the KS test 
     measures if the two samples were drawn from 
     different distributions and does not make that distinction.}
     \label{fig:dm-simulation}
\end{figure*}

\clearpage

\begin{figure*}
     \centering
     \includegraphics[width=\textwidth]{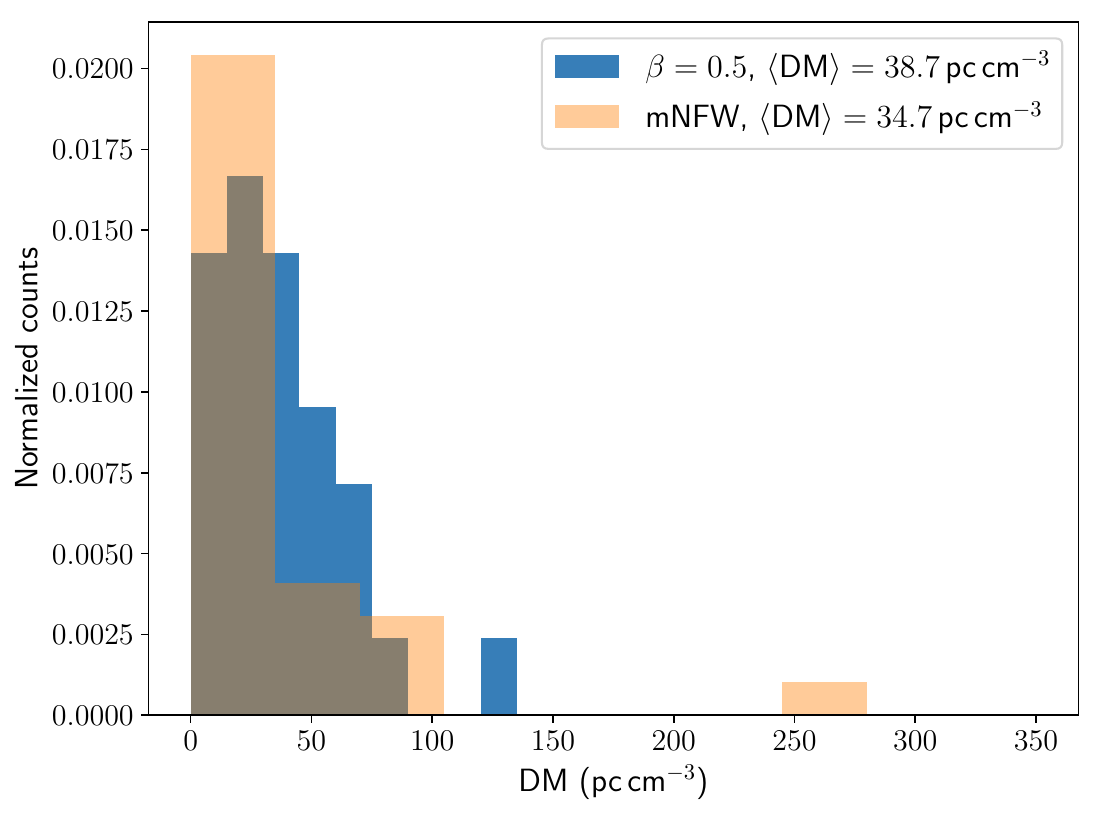}
     \captionsetup{labelformat=empty}
     \caption{{\bf Extended Data Figure 2: Predicted DM excesses for CHIME/FRB galaxy intersections assuming isolated galaxy halos.} The histograms show the relative binned counts of DMs accrued by the fiducial sample of 26 FRBs with $b_{\perp}<200$\,kpc. Two models for the radial density distribution of the CGM are shown: a `beta' model with $\beta=0.5$, and a modified NFW model. Results are shown in blue and orange respectively. The mean DMs of the data shown in each histogram are noted in the figure legend.}
     \label{fig:pred_hist}
\end{figure*}

\clearpage

\begin{figure*}
    \centering
    \includegraphics[width=0.7\textwidth]{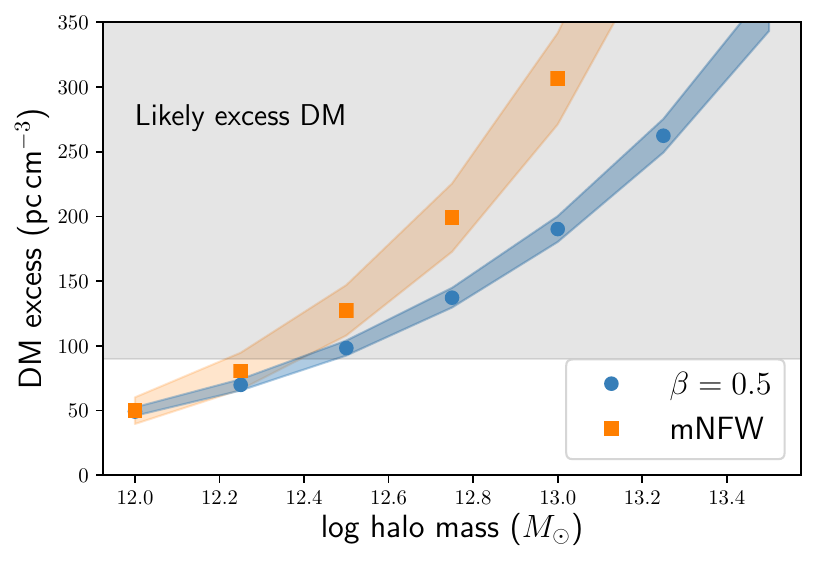}
    \includegraphics[width=0.7\textwidth]{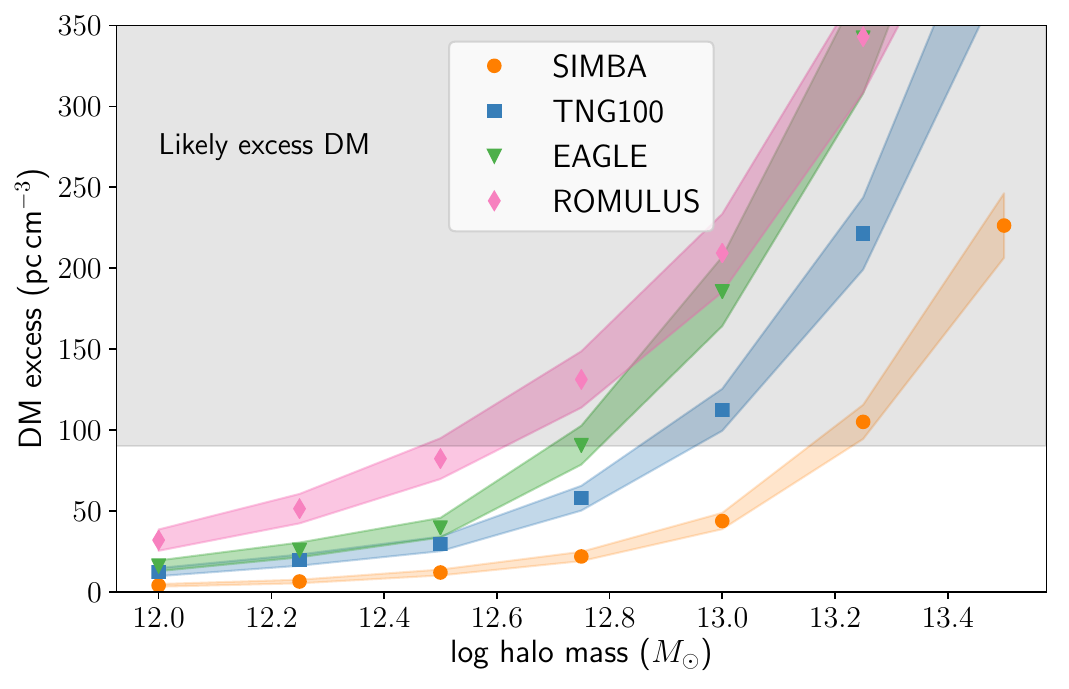}
    \captionsetup{labelformat=empty}
    \caption{{\bf Extended Data Figure 3: predicted DM excesses for 26 FRBs intersecting halos of different masses.} \textit{Top:} We use a `beta' model with $\beta=0.5$ and a modified NFW model for the radial density distribution. Results are shown in blue and orange respectively. For each halo mass, $1\sigma$ error ranges are shown based on 100 simulations of samples of 26 FRB intercepts within the virial radii. The FRB positions relative to the halo centres are simulated using the offsets and CHIME/FRB Catalog 1 position uncertainties of the 26 FRBs in Table 1. The grey shaded area indicates the likely (95\% confidence) DM excess. $f_{\rm gas}=1$ is assumed. \textit{Bottom:} Same as top, but with values of $f_{\rm gas}$ specific to each halo mass derived from four cosmological simulations (see text for details). A modified NFW radial-density model is assumed.}
    \label{fig:pred_offset}
\end{figure*}

\end{document}